\documentclass[twocolumn,twoside,slac_two]{revtex4}
\usepackage{graphicx}
\usepackage{fancyhdr}
\newcommand{\hess}{\textsc{H.E.S.S.}}

\newcommand{\fer}{{\sl {\it Fermi}}}
\newcommand{\fla}{\fer-LAT}

\newcommand{\grs}{$\gamma$-rays}

\newcommand{\pg}{PG~1553+113}

\newcommand{\bestz}{$z=0.49\pm0.04$}

\begin{document}

\title{Probe of Lorentz Invariance Violation effects and determination of the distance of \pg.}

%

\author{D.~A.~Sanchez}
\affiliation{Laboratoire d'Annecy-le-Vieux de Physique des Particules, Universit\'e de Savoie, CNRS/IN2P3}

\author{F.~Brun}
\affiliation{DSM/Irfu, CEA Saclay}

\author{C~Couturier, A~Jacholkowska, J.-P.~Lenain}
\affiliation{LPNHE, Universit\'e Pierre et Marie Curie Paris 6, Universit\'e Denis Diderot Paris 7, CNRS/IN2P3}

\author{On behalf of the \fer\ and \hess\ collaborations}

\begin{abstract}
The high frequency peaked BL Lac object \pg\ underwent a flaring event in 2012. The High Energy Stereoscopic System (H.E.S.S.) observed this source for two consecutive nights at very high energies (VHE, $E>$100~GeV). The data show an increase of a factor of three of the flux with respect to archival measurements with the same instrument and hints of intra-night variability. The data set has been used to put constraints on possible Lorentz invariance violation (LIV), manifesting itself as an energy dependence of the velocity of light in vacuum, and to set limits on the energy scale at which Quantum Gravity effects causing LIV may arise. With a new method to combine H.E.S.S. and \textit{Fermi} large area telescope data, the previously poorly known redshift of PG 1555+113  has been determined to be close to the value derived from optical measurements.

\end{abstract}

\maketitle

\thispagestyle{fancy}

\section{Introduction}
\pg\ is a high frequency peaked BL Lac object located in the Serpens Caput constellation. The object has been detected in VHE by \hess\ \cite{hess1553} in 2006 and in high energies (HE, 100~MeV$<$E$<$300~GeV) by \fer\ \cite{fermi}. The $\gamma$-ray spectrum presents the largest HE-VHE spectral break measured to date \cite{fermi,sanchez}. The source has an unknown redshift despite several attempts to measure it. The best estimate to-date, made by spectroscopy \cite{dan}, is 0.43$<z<$0.58.

In April 2012, the source underwent a flare reported by the MAGIC collaboration \cite{atel}. Subsequently, the source has been observed by the \hess\ telescopes. The data are used in this work to constrain its redshift and to probe a possible Lorentz invariance violation.

\section{Data Analysis}
\subsection{H.E.S.S. data analysis}
The \hess\ telescopes have observed \pg\ in April 2012 for two nights. Data were analysed using the {\tt Model} analysis \cite{model} with {\tt Loose} cuts. The object has been detected with a significance of 21.5$\sigma$ in 3.5 hours of live time. The source spectrum is well fitted by a power-law model of the form: 
$$F(E) \propto E_*^{-(4.85\pm 0.25)}$$
where $E_*=E/(327\ {\rm GeV})$ and the flux is found to be 3.5 times higher than the measurements made in 2005-2006 \cite{hess1553}.

Indications of intra-night variability have been found with the fit to a constant of the light-curve yielding a $\chi^2$ of $21.34$ for $7$ d.o.f. ($P_{\rm \chi^2}=3.3\times 10^{-3}$).

Data taken in 2005-2006 were re-analysed and the spectrum is, in this case, well fitted by a log-parabola model:
$$F(E) \propto E_*^{-(5.39\pm0.43) -(3.95\pm1.40)~log_{10}E_*}$$
where $E_*=E/(360\ {\rm GeV})$. Both spectra (archival and flare) are presented on figure \ref{SED}.

\subsection{\fla\ data analysis}

The \fla\ data, from 300~MeV to 300~GeV, have been analysed with the ScienceTools {\tt V9R32P5} and instrumental response functions {\tt P7REP\_SOURCE\_V15}. A region of interest of 15 degrees has been used and the sky model has been built using the third \textit{Fermi} catalog \cite{3FGL}.

\begin{figure}
\includegraphics[width=85mm]{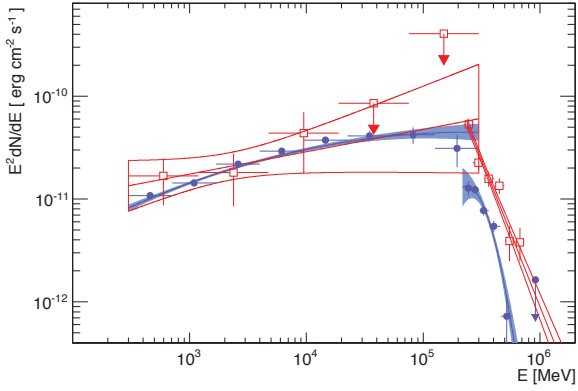}
\caption{Spectral energy distribution in $gamma$-ray measured with \fla\ and H.E.S.S. during the flare (red) and in the pre-flare state (see text) in blue.}
\label{SED}
\end{figure}

Data contemporaneous to the \hess\ exposures taken in 2012 are well fitted by a power-law of index $\Gamma= 1.72\pm 0.26$. The pre-flare data are defined by the data taken from August 8, 2008 up to March 1st 2012. The measured spectrum is described by a log-parabola model (Fig. \ref{SED}).

Variability has been probed before, during and after the flare using a bayesian blocks analysis \cite{bb}. No counterpart to the VHE flare was found in the HE light curve.

\section{Determination of the redshift}

The extragalactic background light (EBL) is a field of infrared photons that interacts with the VHE \grs\ on their way to Earth. This absorption leaves a footprint on the source spectrum that is used in this work to constrain its distance using a new bayesian model. The Bayes theorem reads $P(\Theta|Y) \propto P(\Theta)  P(Y|\Theta)$ where $Y$ stands for the data and $\Theta$ the parameters. The likelihood, $P(Y|\Theta)$, is minimized during the spectrum determination. The model used here is a power-law corrected for EBL absorption i.e $ \phi = N\times (E/E_0)^{-\Gamma}\times e^{-\tau(E,z)}.$ and then $\Theta$ is $N$, $\Gamma$ and $z$.

To construct the prior $ P(\Theta)$, the following assumptions have been made:
\begin{itemize}
\item EBL-corrected power-law cannot be harder than the \fer\ measurement, the prior being then a truncated Gaussian of mean $\Gamma_{\fer}$ and width $\sigma$ that accounts for statistical and systematic uncertainties of both instruments.
\item Softening of this power-law, that arises from emission effects, is permitted with a constant probability. 
\item It is also assumed that distant sources are harder to detect and that $P(z) \propto exp(-\tau(z))$.
\end{itemize}

\begin{figure*}[Ht!]
     \includegraphics[width=135mm]{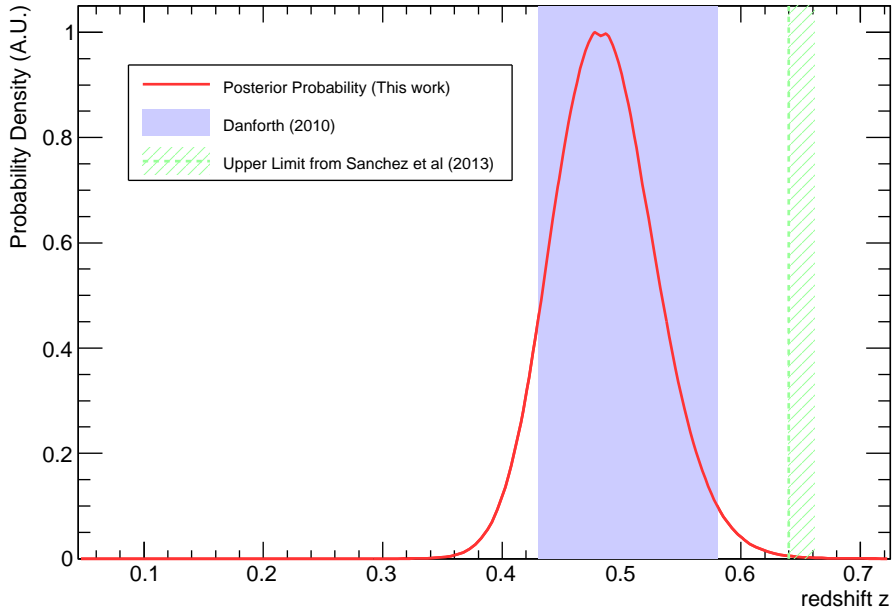}
\caption{Posterior probability for the redshift determination and comparaison with other measurements.}
\label{Prob}
 \end{figure*}

The EBL absorption is computed using the model of \cite{fra}. Marginalising over the parameters gives a redshift \bestz\ (Fig. \ref{Prob}) in agreement with other measurement \cite{dan} or constraints derived using GeV-TeV data \cite{sanchez}.

\section{Lorentz invariance violation (LIV)}
Tests for a possible LIV effect were performed by searching for a non-zero dispersion parameter $\tau_n (\sim \frac{\Delta t}{(\Delta E)^n})$ in the H.E.S.S. data of the flare. This is done by testing a correlation between arrival times of the photons and their energies.

A maximum likelihood analysis based on \cite{liv1}, has been modified to tackle the non-negligible background present in the data.

For $n_{\rm ON}$ events recorded in the ON-source region with arrival times $t_i$ and energies $E_i$, the likelihood reads:

$\mathit{L}(\tau_n) = \prod_{i = 1}^{n_{\rm ON}} P(E_i, t_i|\tau_n)$ with:
\vskip -.3cm
$$P(E_i, t_i|\tau_n)  
= w_s \cdot P_{\textrm{\tiny Sig}}(E_i, t_i|\tau_n) + (1-w_s) \cdot P_{\textrm{\tiny Bkg}}(E_i, t_i)$$
\vskip -.3cm

The probability $P_{\textrm{\tiny Sig}}$ was mainly determined from a parametrization of the light curve at low energies parametrization while $P_{\textrm{\tiny Bkg}}$ was built assuming a constant background. The factor $w_s$ accounts for the relative weight of signal events with respect to background events.

Constraints on $\tau_n$ led to lower limits on the Quantum Gravity energy scale $E_{\rm QG}$.
The 95\% 1-sided lower limits for the subluminal case are:
$\textrm{E}_{\rm QG,1}>4.32\times 10^{17}$~GeV and $\textrm{E}_{\rm
QG,2}>2.11\times 10^{10}$~GeV for linear and quadratic LIV effects,
respectively. Figure 3 compares these results with other limits from less distant AGN flares. While the statistics is more limited here, the distance of the source makes the sensitivity to possible LIV effects comparable to previous results.

\begin{figure}[H]
  \begin{minipage}{\textwidth}
  
     \includegraphics[width=80mm]{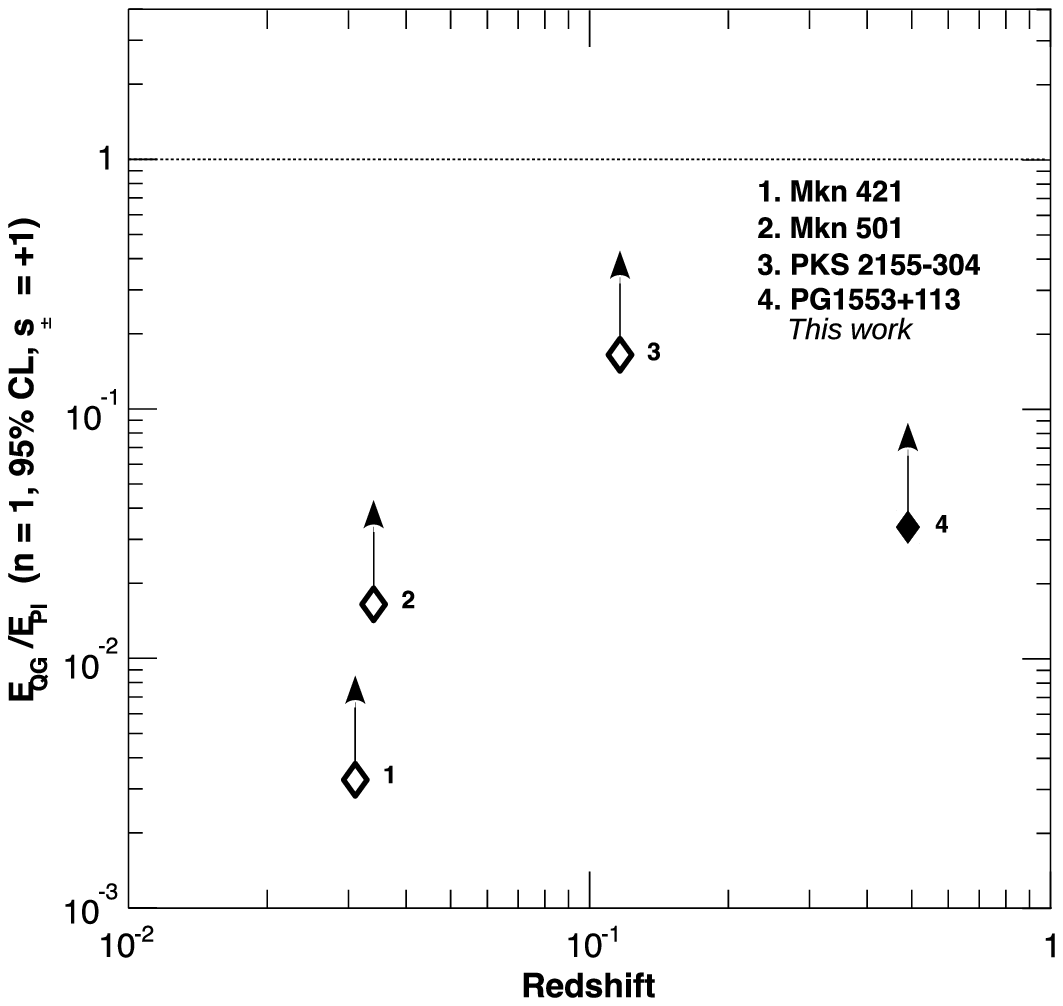}
     \includegraphics[width=80mm]{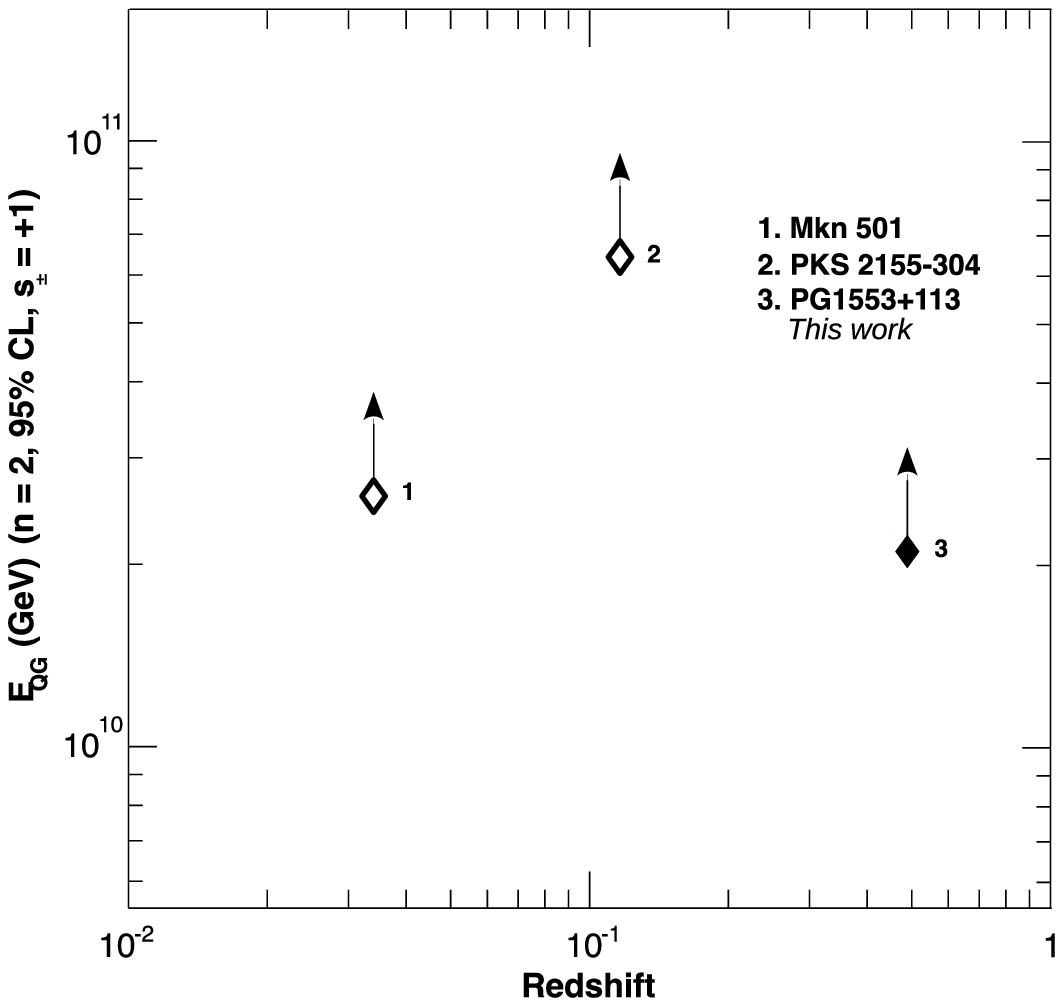}
\caption{ Lower limits on $\textrm{E}_{\rm QG,1}$ from linear dispersion (left) and on $\textrm{E}_{\rm QG,2}$ from quadratic dispersion (right) for the subluminal case obtained with AGN as a function of redshift.}
\end{minipage}
\label{liv}
 \end{figure}

\section{Conclusions}
 The VHE emitter \pg\ underwent a flaring event in VHE with an increase of its flux by a factor of 3.5. No counterpart of this flare was found in the HE regime by \fer. 

This data set has been used to constrain the redshift of the source to be \bestz\ using a novel method based on $\gamma$-ray data. The flare is also used to put lower limits on the LIV effect with $\textrm{E}_{\rm QG,1}>4.32\times 10^{17}$~GeV and $\textrm{E}_{\rm
QG,2}>2.11\times 10^{10}$~GeV for linear and quadratic effects.

\bigskip 

\section*{Acknowledgments} \label{Ack}

The support of the Namibian authorities and of the University of Namibia in facilitating the construction and operation of H.E.S.S. is gratefully acknowledged, as is the support by the German Ministry for Education and Research
(BMBF), the Max Planck Society, the French Ministry for Research, the CNRS-IN2P3 and the Astroparticle Interdisciplinary Programme of the CNRS, the U.K. Science and Technology Facilities Council (STFC), the
IPNP of the Charles University, the Polish Ministry of Science and Higher Education, the South African Department of Science and Technology and National Research Foundation, and by the University of Namibia. We appreciate
the excellent work of the technical support staff in Berlin, Durham, Hamburg, Heidelberg, Palaiseau, Paris, Saclay, and in Namibia in the construction and operation of the equipment.

The \textit{Fermi}-LAT Collaboration acknowledges support for LAT development, operation and data analysis from NASA and DOE (United States), CEA/Irfu and IN2P3/CNRS (France), ASI and INFN (Italy), MEXT, KEK, and JAXA (Japan), and the K.A.~Wallenberg Foundation, the Swedish Research Council and the National Space Board (Sweden). Science analysis support in the operations phase from INAF (Italy) and CNES (France) is also gratefully acknowledged.

The work of DS has been supported by the Investissements d'avenir, Labex ENIGMASS.

\bigskip 

\end{document}